# Meditative absorption shifts brain dynamics toward criticality

Jonas Mago[1*], Joshua Brahinsky[2], Mark Miller[3], Charlotte Maschke[1], Heleen A. Slagter[4], Shaila Catherine[5], Ruben E. Laukkonen[6], B. Rael Cahn[7], Matthew D. Sacchet[8], Wangmo Dixey[9], Richard Dixey[10], Soham Rej[2,11], Michael Lifshitz[2, 11]

[1] Integrated Program in Neuroscience, McGill University, Montréal, Quebec, Canada
[2] Division of Social and Transcultural Psychiatry, McGill University, Montréal, Quebec, Canada
[3] Monash Center for Consciousness and Contemplative Studies, Monash University
[4] Department of Experimental and Applied Psychology & Institute Brain and Behavior Amsterdam, Vrije Universiteit Amsterdam
[5] Meditation Teacher, San Jose, California, USA
[6]Faculty of health, Southern Cross University, Gold Coast, Australia
[7]Department of Psychiatry & Brain and Creativity Institute, University of Southern California, Los Angeles CA
[8]Meditation Research Program, Department of Psychiatry, Massachusetts General Hospital, Harvard Medical School, Boston, MA, USA
[9] Dharma College, Berkeley, California
[10] Institute for the Study of Extraordinary States of Mind (ISEOM), Berkeley, California
[11] Lady Davis Institute for Medical Research, Jewish General Hospital, Montréal, Quebec, Canada

* Corresponding author: jonas.h.mago@gmail.com

**Abstract**

Criticality describes a regime between order and chaos that supports flexible yet stable information processing. Here we examine whether neural dynamics can be volitionally shifted toward criticality through the self-regulation of attention. We examined ten experienced practitioners of meditation during a 10-day retreat, comparing refined states of meditative absorption, called the "jhānas", to regular mindfulness of breathing. We collected electroencephalography (EEG) and physiological data during these practices and quantified the signal's dynamical properties using Lempel–Ziv complexity, signal entropy, chaoticity and long-range temporal correlations. In addition, we estimated perturbational sensitivity using a global auditory oddball mismatch negativity (MMN) during meditation. Relative to mindfulness, jhāna was associated with pronounced self-reported sensory fading, slower respiration, higher neural signal diversity across multiple measures, reduced chaoticity, and enhanced MMN amplitude over frontocentral sites. Spectral analyses showed a flatter aperiodic 1/f component and a frequency-specific reorganization of long-range temporal correlations. Together, increased diversity with reduced chaoticity and heightened deviance detection indicate a shift toward a metastable, near-critical regime during jhāna. We propose an overlap of the phenomenology of jhāna with minimal phenomenal experiences in terms of progressive attenuation of sensory content with preserved tonic alertness. Accordingly, our findings suggest that criticality is a candidate neurophysiological marker of the absorptive, minimal-content dimension of the minimal phenomenal experience.

## Introduction

The human brain is a complex dynamical system that must continually balance stability with flexibility to facilitate perception and action (Friston et al., 2012). Recent work in computational neuroscience has proposed the concept of criticality—a state poised between order and chaos—as a requirement for healthy brain function and the emergence of higher cognitive capacities, including consciousness (Gervais et al., 2023; Kim & Lee, 2019; Maschke et al., 2024; O'Byrne & Jerbi, 2022; Shew & Plenz, 2013; Tagliazucchi, 2017; Walter & Hinterberger, 2022; Wilting & Priesemann, 2019). Poised at criticality, the brain is thought to maximize information integration, computational capability and sensitivity to perturbation (Breyton et al., 2024; Jobst et al., 2021; Perl et al., 2022). Evidence shows that ingesting psychedelic substances can bring the brain closer to criticality while anesthesia moves the brain further away (Carhart-Harris, 2018; Gervais et al., 2023; Jobst et al., 2021; Krystal et al., 1996; Ort et al., 2023; Toker et al., 2022, 2024; Varley et al., 2020). In this study, we test whether it is possible to volitionally shift neural dynamics toward criticality by self-regulating the state of attention, without pharmacological interventions.

Brain criticality has been characterized empirically in several ways, including avalanche statistics, long-range temporal correlations, and critical slowing (Dawes & Freeland, 2008; Gervais et al., 2023; Gottwald & Melbourne, 2009; Kanders et al., 2017; O'Byrne & Jerbi, 2022; Steeb & Andrieu, 2005; Wilting & Priesemann, 2019). Among other types of criticality (see O'Byrne & Jerbi, 2022 as a review), the closeness to the edge-of-chaos is one way to approximate criticality from brain signals through the joint assessment of signal complexity and chaoticity (Steeb & Andrieu, 2005; Toker et al., 2022, 2024). In this framework, a system closer to criticality exhibits a richer repertoire of neural states (i.e., higher signal diversity) while avoiding runaway instability (i.e., chaoticity) (Maschke et al., 2024; Steeb & Andrieu, 2005; Toker et al., 2024). In addition to measuring signal properties, criticality can be assessed by measuring how a system responds to external perturbation, for example using the auditory mismatch negativity (MMN). The MMN is an event-related brain response elicited by an unexpected auditory stimulus in a sequence of repeating auditory stimuli (Näätänen et al., 2007). The MMN has been proposed as a perturbational marker of criticality: weaker responses suggest a move away from critical dynamics, while stronger responses signal proximity to criticality (Shi et al., 2022). A system closer to criticality should then exhibit an increase in signal diversity, a decrease of chaoticity, and a stronger MMN.

Previous studies suggest that meditation can shift brain dynamics closer to criticality, although the evidence is mixed. First, a recent review by Atad et al. (2025) suggests that meditation increases signal diversity, the first requirement of brain criticality (Do et al., 2023; Irrmischer et al., 2018; Kakumanu et al., 2018; Lu & Rodriguez-Larios, 2022; Pascarella et al., 2025). Yet, other studies found decreased signal diversity (e.g., Aftanas & Golocheikine, 2002; Gupta et al., 2021; Kakumanu et al., 2018; Young et al., 2021). Second, a system closer to criticality does not only exhibit a richer repertoire of neural states (i.e., higher signal diversity), but also avoids runaway instability (i.e., chaoticity). One study by Gao et al., (2016) found that mindfulness of breathing, as taught in the popular Mindfulness-Based Stress Reduction program, reduced chaoticity as measured by wavelet entropy of EEG and heart rate signals compared to resting. However, the evidence remains sparse. Third, findings on the relationship between meditation and the MMN are divided. Early studies reported an enhanced MMN after focused attention meditation (Biedermann et al., 2016; Fucci et al., 2018; Srinivasan & Baijal, 2007). However, more recently Fucci et al. (2022) found that neither focused attention nor open presence meditation impacted the MMN compared to a control condition (silent movie) in either experienced practitioners or novices. The inconsistencies in these findings may be due to differences in practices, limited sample sizes, uneven metrics, and mismatched control conditions. Despite some suggestive evidence, this mixed bag of results leaves open the question of whether meditation reliably moves the brain toward criticality.

Meditative practices differ in the degree to which they cultivate *absorption*—a progressive focusing of attention in which awareness becomes increasingly unified and disengaged from ordinary sensory and cognitive contents. While some practices, such as open monitoring styles of meditation, emphasize broad awareness of ongoing sensory and mental events, others explicitly aim at cultivating deep absorption. In these practices, attention narrows and stabilizes on a chosen object until peripheral contents fade, producing states marked by stillness, unification, and sensory withdrawal. We propose that specifically isolating the process of meditative absorption from the other dimensions of meditative practice—such as meta-cognition and attention regulation—may help to clarify the mixed findings on meditation and brain criticality. We situate absorptive depth within the framework of minimal phenomenal experience (MPE), which formalizes states where phenomenal content is progressively reduced or absent, yet basic, tonic awareness is preserved (Metzinger, 2020, 2024). In minimizing the interference of sensory and cognitive content, yet maintaining a globally responsive system, MPE may characterize a balance between stability and flexibility that could link deep absorption to criticality.

In our extended phenomenological interviews with experienced practitioners of *jhāna*—a Theravāda Buddhist practice that cultivates progressively deeper states of absorption—they frequently describe emerging from deep absorption with a sense of freshness, openness, and heightened mental flexibility, qualities that echo the dynamical hallmarks of a system operating near criticality. *Jhāna* refers to a sequence of states in which attention deepens to the point that experience is said to become radically simplified and unified. The first four states of absorption are called "fine-material jhānas" and are the subject of the current study. After a few days on retreat, the expert participants we studied were able to access these states of pronounced sensory fading with relative reliability and stability (Yang, Chowdhury, et al., 2024; Yang, Sparby, et al., 2024).

To isolate the effects of meditative absorption, we compared jhāna to mindfulness of breathing, a common meditative practice that was familiar to all our participants. Mindfulness of breathing practice involves attending to the ongoing flux of breathing-related sensations in the body, but without the progressive absorption of attention into a mental object that is characteristic of jhāna. While the capacity of mindfulness (sati, or recollecting the mind) is also central to the development of jhāna, the attention is placed differently in each practice. In jhāna, attention is placed on a "mental" sign, a subtle mental representation of the breath called the breath nimitta. The nimitta is described as an internal light or mental image that arises when attention becomes highly stabilized on the breath. In jhāna, sustained focus on the nimitta draws attention inward, producing deep absorption and a sense of unification of the field of awareness (Catherine, 2008, 2011; Sayadaw, 1995; Snyder & Rasmussen, 2009). Given these similarities and differences, we employed mindfulness of breathing as a control condition to isolate the effect of absorption on criticality; like jhana, mindfulness of breathing is anchored in paying attention to the breath, but unlike jhana, it does not involve becoming absorbed in the breath to the exclusion of the rest of experience.

Although recent studies have begun to examine the neurophysiological correlates of jhāna (Chowdhury et al., 2025; DeLosAngeles et al., 2016; Demir et al., 2025; Dennison, 2019; Hagerty et al., 2013; Potash, van Mil, et al., 2025; Potash, Yang, et al., 2025; Sparby & Sacchet, 2024; Yang, Chowdhury, et al., 2024; Yang, Sparby, et al., 2024, 2024), most have compared jhāna to resting baselines, making it difficult to distinguish effects specific to jhāna from those common to meditative practice more generally, such as relaxation and changes in brain spectral power (Lutz et al., 2004). Moreover, prior research was not conducted in retreat settings, which many of our participants said would be important to support the stabilization of jhāna, and other studies based their findings on a single participant (Potash, Yang, et al., 2025; Yang, Chowdhury, et al., 2024; Yang, Sparby, et al., 2024).

The present study is the first to directly compare jhāna meditation to another active form of meditation, thereby isolating the role of absorptive depth in shifting brain dynamics toward criticality. We specifically focused on the practice methodology taught within the Pa-Auk Sayadaw lineage, a tradition known for its rigorous approach to meditative absorption. We recruited ten lay practitioners from North America under the instruction of Shaila Catherine, a senior American Buddhist teacher who trained for several years under Pa-Auk Sayadaw. Data were collected during a 10-day silent residential retreat that provided environmental and attentional support aimed at stabilizing absorption. Across four days, participants engaged in both jhāna and mindfulness meditation (order was counterbalanced) while EEG and physiological signals were continuously recorded. This design enabled us to isolate the neural dynamics specifically associated with absorptive depth. We predicted that, compared to mindfulness of breathing, jhāna's progression toward minimal phenomenal experience would shift brain dynamics closer to criticality, manifesting as increased signal diversity, reduced chaotic dynamics.

In addition, we assessed the impact of jhāna absorption on the brain’s perturbational sensitivity as measured by the auditory MMN. Although systems close to criticality are generally expected to be highly sensitive to perturbation (Breyton et al., 2024; Jobst et al., 2021; Perl et al., 2022), the jhānas are also characterized by profound stability and sensory withdrawal, a state seemingly impervious to disturbance. To assess whether phenomenological stability corresponds with neural stability or, instead, with the heightened sensitivity expected near criticality, we employed an auditory mismatch-negativity (MMN) paradigm: participants meditated while hearing repetitive sequences of five vowels, with occasional deviants that violated the pattern. Prior research shows that these deviations trigger an MMN, greater negativity over frontocentral scalp regions, even when participants are engaged in another task (e.g., reading) (Garrido et al., 2009; Näätänen et al., 2007). The MMN thus provides a measure of the brain's automatic response to unexpected sensory perturbations. If jhāna achieves its phenomenological withdrawal through sensory decoupling, MMN responses should be attenuated. Conversely, if jhāna represents a critical brain state, the heightened sensitivity to unexpected perturbation should manifest as enhanced MMN amplitude; revealing that states of deep absorption have an experiential stillness while external responsiveness is facilitated.

Notably, recent studies show that the MMN response disappears in the absence of consciousness, as in deep sleep (Strauss et al., 2015), but may remain present during an advanced, objectless meditation state called open presence meditation (Fucci et al., 2018, 2022). However, in contrast to the standard auditory tone, the deviant auditory tone in these studies was not only unexpected but also physically distinct from the preceding tone. The observed MMN during open presence meditation may in part reflect preserved passive sensory response adaptation, i.e., activity by fresh afferent neural populations activated by deviant stimuli physically distinct from the directly preceding stimuli (Garrido et al., 2009; May & Tiitinen, 2010), rather than detection of environmental deviancy. To address this potential confound, our deviant vowel was unexpected based on the preceding sequence, yet physically identical to the preceding vowel (e.g., "AOAO**O**"/ "OAOA**A**" vs. "AOAO**A**"/ "OAOA**O**"). This way, sensory adaptation would affect ERP activity in the opposite direction from prediction error signaling, if at all, as contributions from sensory adaptation are minimized with a single vowel repetition (Strauss et al., 2015). We initially predicted a reduced MMN response, mistakenly reasoning that the pronounced sensory fading characteristic of jhāna would outweigh an increase in perturbational sensitivity.

The central aim of this study is to test whether brain dynamics can be volitionally shifted closer to criticality. As a unique test case, we examined experienced practitioners of jhāna meditation, a contemplative state characterized by reports of profound absorption and sensory fading. We hypothesized that jhāna meditation would be associated with greater criticality of brain dynamics compared to a non-absorptive practice of mindfulness of breathing, and that the depth of absorption would positively correlate with shifts in brain dynamics toward criticality.

Hypotheses and analyses were preregistered (https://osf.io/z9ycg).

## Results

### *Jhāna induces sensory fading and autonomic reorganization*

Ten experienced jhāna meditators (mean age = 61.3 years; 7 female; average of 8,820 lifetime meditation hours, SD = 9,090; range = 2,260hr to 32,173hr) participated in a within-subject design comparing daily 40-minute sessions of jhāna and mindfulness of breathing across four consecutive days of a residential meditation retreat. Each participant completed both conditions in a counterbalanced order while EEG and physiological signals were continuously recorded.

Subjective and physiological data confirmed that jhāna induces a distinct meditative state marked by reduced sensory engagement and altered bodily rhythms. Compared to mindfulness, participants reported a significantly greater fading of sensory experiences during jhāna meditation ($p = 0.014$, $\beta = 0.41$, $R^2 = 0.54$); with linear mixed-effects modeling revealing a progressive increase across the first four jhāna stages ($p < 0.001$); consistent with canonical descriptions of jhāna as a state of deep absorption and withdrawal from sensory input. The fading of sensory experience was not significantly influenced by the order of the meditation blocks within a session ($p=0.995$) but deepened across the four consecutive measuring days ($p=0.002$). The self-reported sensory fading furthermore highly correlated with the self-reported stability of the current meditation state ($p=0.002$).

Physiological markers complemented these self-reports, revealing a shift toward slower and more stable autonomic patterns. Respiration rate was significantly lower during jhāna ($p = 0.016$, $\beta = -0.22$, $R^2 = 0.86$), reflecting a transition to slower-paced, absorption-aligned breathing. In parallel, the complexity of the respiration signal, quantified via Lempel–Ziv complexity, was significantly higher ($p < 0.001$, $\beta = 0.33$, $R^2 = 0.74$), indicating more flexible respiratory dynamics. In contrast, cardiac and electrodermal measures showed no significant differences in mean heart rate ($p = 0.247$), heart rate variability ($p = 0.407$), and skin conductance ($p = 0.263$).

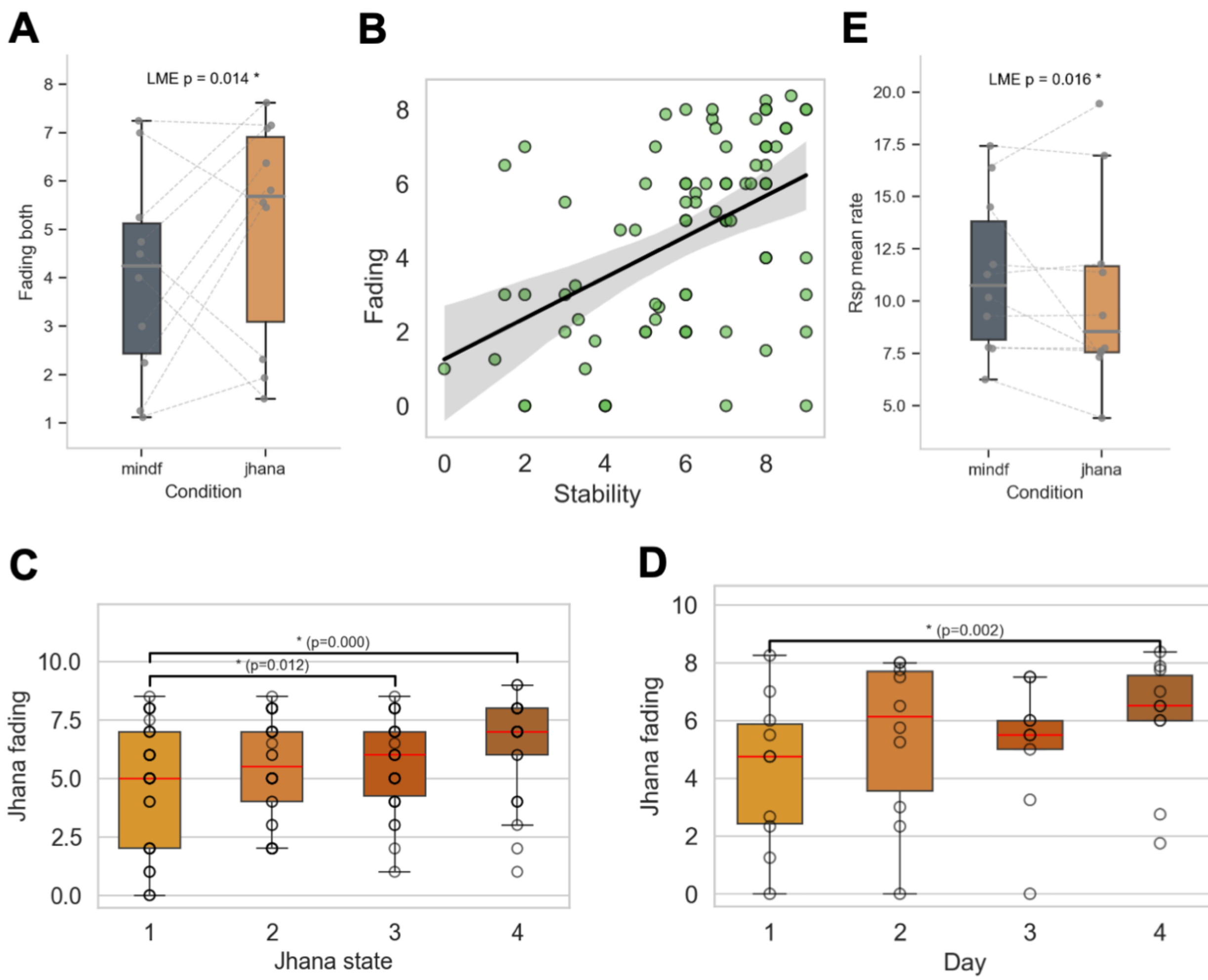

**Figure 1: Jhāna induces progressive sensory fading and autonomic reorganization.** (A) self-reports show that there is significantly more sensory fading during jhāna than mindfulness ($p = 0.014$, $\beta = 0.41$, $R^2 = 0.54$); (B), the self-reports of sensory fading correlate with self-reports of the stability of the meditation ($p=0.002$), and (C) self-reports of sensory fading significantly deepened across the four consecutive jhāna states ($p<0.000$) as well as (D) across the four consecutive days of data collection ($p=0.002$). (E) We also found a significant slowing of breathing rates during jhāna compared to mindfulness ($p = 0.018$, $\beta = -0.22$, $R^2 = 0.5$).

***Jhāna increases signal complexity compared to mindfulness***

In line with our preregistered hypothesis, we find that jhāna meditation significantly increased Lempel-Ziv Complexity (LZC) compared to mindfulness ($p = 0.004$, $\beta = 0.53$, $R^2 = 0.45$). LZC is a widely used measure of signal diversity that quantifies the compressibility of a binary time series. This makes LZC a sensitive index of signal diversity, known to increase with psychedelics, and to decrease under anesthesia and sleep compared to wakefulness (Sarasso et al., 2021; M. Schartner et al., 2015; M. M. Schartner et al., 2017; Toker et al., 2022). LZC is only one of many possible measures for signal complexity (Makowski et al., 2021), we therefore computed three complementary diversity measures: sample entropy, permutation entropy, hjorth mobility, and the aperiodic component.

Unlike LZC, which focuses on pattern novelty, sample entropy captures temporal unpredictability—how hard it is to forecast the future from the recent past. Hjorth mobility, on the other hand, estimates the spectral variability of the signal by comparing its first derivative to the original signal, offering a frequency-domain perspective on complexity. While each of these metrics highlights a different aspect of diversity—pattern richness (LZC), temporal irregularity (sample entropy), broadband spectral balance (spectral slope), and frequency dispersion (mobility)—all showed consistent effects. Compared to mindfulness, Jhāna shows increased levels of sample entropy ($p < 0.001$, $\beta = 0.61$, $R^2 = 0.53$), permutation entropy ($p < 0.001$, $\beta = 0.63$, $R^2 = 0.9$) and Hjorth mobility ($p = 0.001$, $\beta = 0.61$, $R^2 = 0.53$), supporting our findings with LZC across multiple dimensions of signal complexity.

We complemented these signal complexity metrics with an analysis of spectral features using the aperiodic component, associated with scale-free dynamics and excitation-inhibition balance. We find that, compared to mindfulness, jhāna shows a flatter 1/f slope ($p < 0.001$, $\beta = 0.57$, $R^2 = 0.66$), a pattern that has previously been interpreted and found to be highly correlated with other measures of signal complexity (Höhn et al., 2024; Medel et al., 2023). In the power spectral analysis, we see a significant decrease in alpha power ($p < 0.001$, $\beta = -0.27$, $R^2 = 0.96$), and a significant increase in gamma power ($p < 0.001$, $\beta = 0.65$, $R^2 = 0.75$). No significant differences were found in delta ($p=0.862$), theta ($p=0.098$), and beta ($p=0.075$) (cf. Lee et al., 2018).

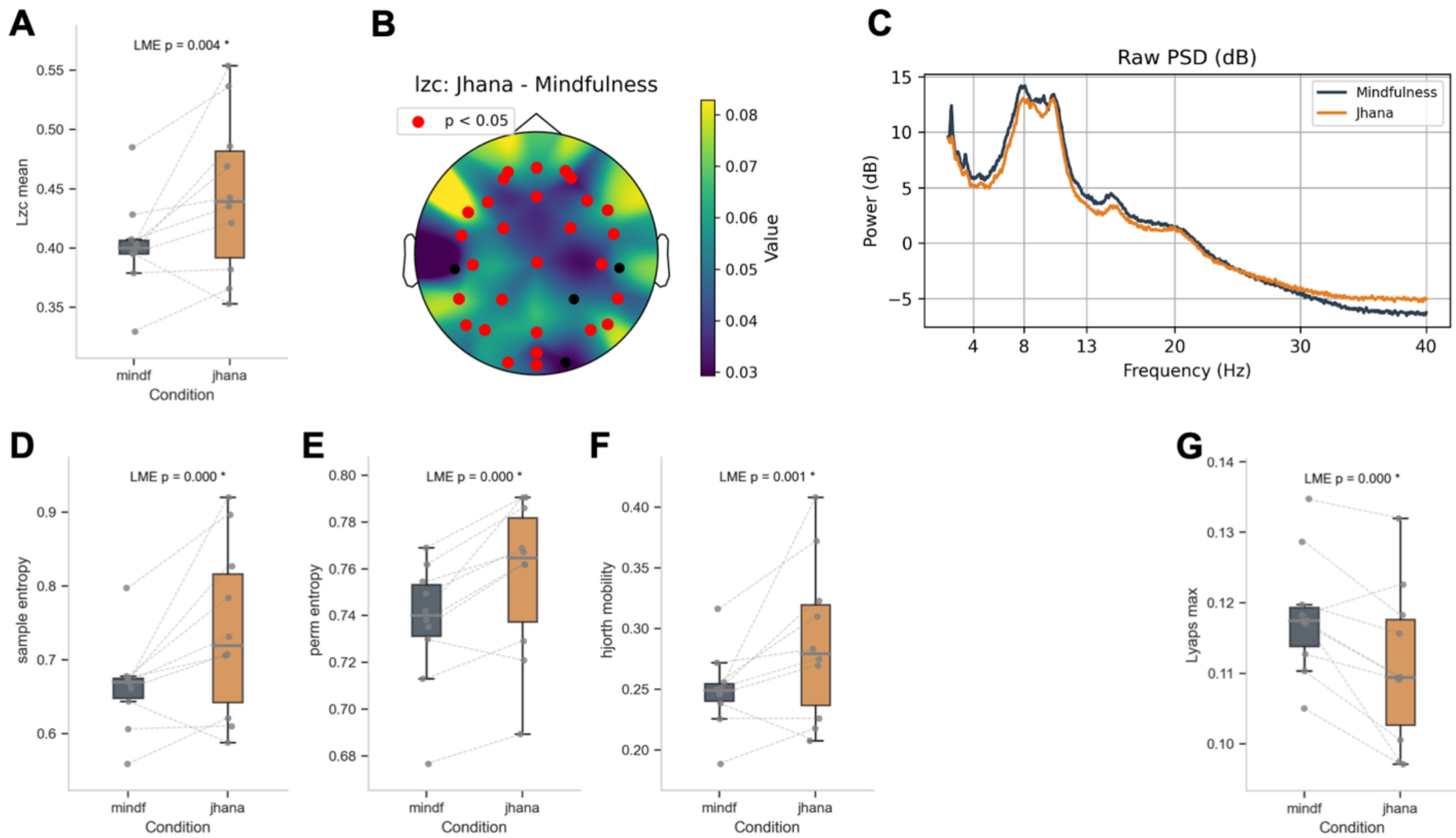

**Figure 2: Jhāna increases neural signal complexity compared to mindfulness.** (A) Lempel-Ziv complexity is significantly higher during jhāna ($p = 0.004$), notably, variability across participants was lower during mindfulness and more pronounced during jhāna. (B) this difference is significant for 28 out of the 32 channel positions (i.e., $p < 0.05$). (C) We confirmed this finding of increased signal diversity by showing similar increases in signal diversity with a flattening of the 1/f curve ($p < 0.001$) as well as significantly higher measures of (D) sample entropy ($p < 0.001$), (E) permutation entropy ($p < 0.001$), (F) Hjorth mobility ($p = 0.001$). (G) In addition to the measures of signal diversity, Jhāna significantly reduces the largest Lyapunov exponent ($p < 0.001$). The Lyapunov exponent remains > 0 in both conditions, indicating that EEG time series exhibit chaotic behavior during both jhāna and mindfulness, but with slower divergence of nearby trajectories during jhāna, suggesting more stable dynamics.

***Jhāna decreases brain chaoticity compared to mindfulness, indicating brain dynamics closer to criticality***

To assess whether the increased signal diversity observed during jhāna reflects structured dynamics indicative of criticality—rather than unstructured noise—we examined the Lyapunov exponent, a metric that captures the average rate at which nearby trajectories in phase space diverge over time. Smaller values reflect reduced dynamical instability and greater predictability in the system's evolution. As such, the Lyapunov exponent is a measure of chaoticity which quantifies the sensitivity and stability of neural trajectories. Compared to mindfulness, jhāna was associated with a significant reduction in the largest Lyapunov exponent ($p < 0.001$., $\beta = -0.63$., $R^2 = 0.79$), indicating decreased chaoticity. This reduction in chaoticity goes hand in hand with the observed increases in signal complexity—suggesting that jhāna shifts brain dynamics toward a regime that is both richly differentiated and stably organized. Together, these findings indicate that jhāna does not merely increase neural diversity, but selectively reorganizes brain activity toward a flexible, metastable regime characteristic of criticality. This interpretation was further supported by strong negative correlations between Lempel–Ziv complexity and the largest Lyapunov exponent ($r = -0.95$, $p < 0.001$), indicating that greater signal diversity during jhāna co-occurred with reduced chaoticity and enhanced dynamical stability.

### ***Jhāna alters long range temporal correlations***

To further probe the temporal organization of neural dynamics, we analyzed long-range temporal correlations (LRTC) using detrended fluctuation analysis (DFA), which quantifies the degree of self-similarity across time in oscillatory fluctuations. While no significant differences were observed when computing the DFA over the whole spectrum (1–45 Hz; $p = 0.443$, $\beta = 0.15$, pseudo-$R^2 = 0.29$), frequency-resolved analyses revealed distinct band-specific effects. Compared to mindfulness, jhāna meditation was associated with significantly reduced LRTC in the theta (4–8 Hz; $p = 0.050$, $\beta = -0.23$, $R^2 = 0.77$), alpha (8–13 Hz; $p = 0.023$, $\beta = -0.22$, $R^2 = 0.85$), and beta (13–30 Hz; $p = 0.001$, $\beta = -0.41$, $R^2 = 0.77$) band. In contrast, LRTC were significantly higher in the low gamma band (30–45 Hz; $p = 0.039$, $\beta = 0.15$, pseudo-$R^2 = 0.91$), and trended higher in the delta band (1–4 Hz; $p = 0.150$, $\beta = 0.26$, pseudo-$R^2 = 0.42$).

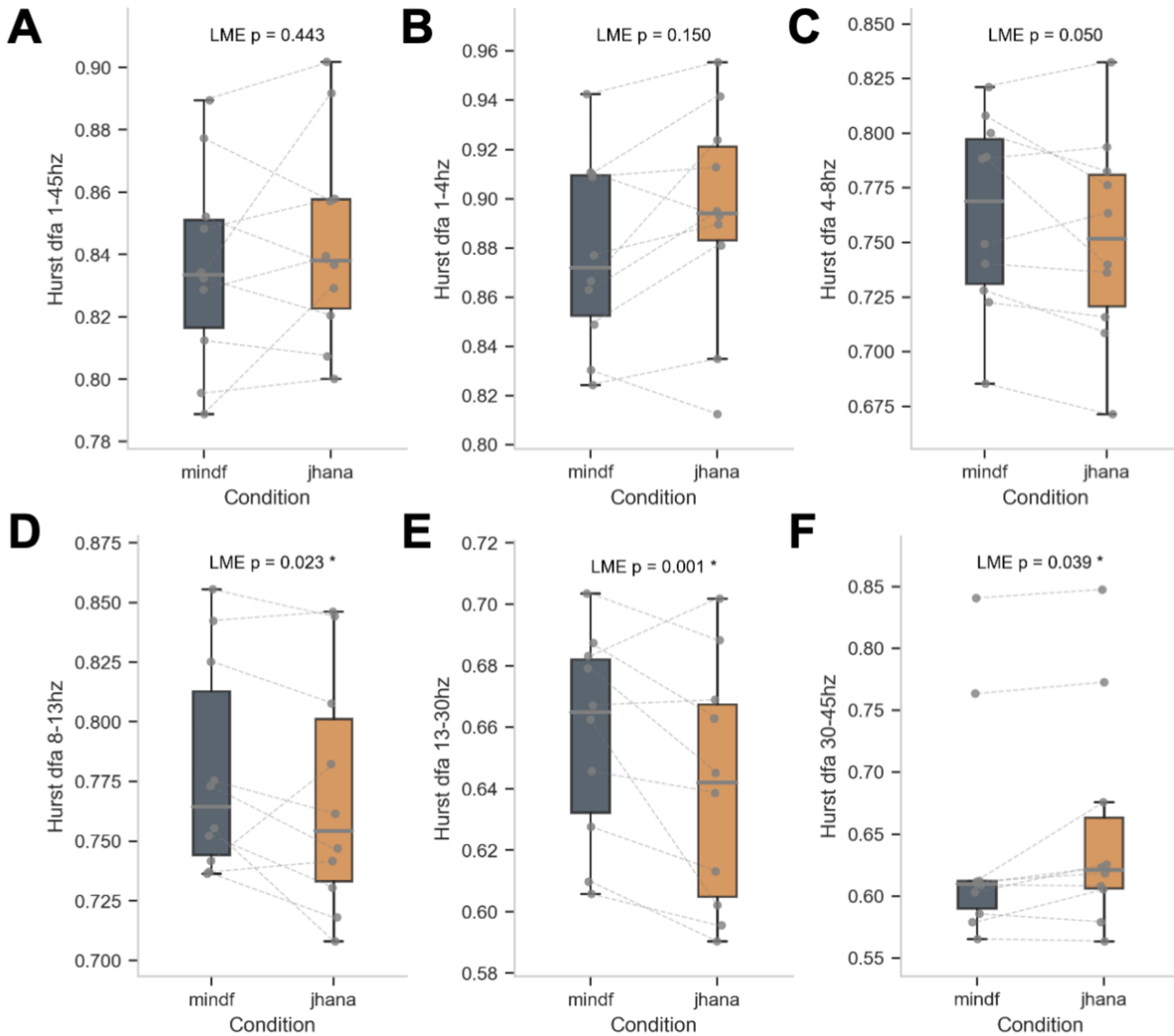

**Figure 3: Frequency-specific reorganization of long-range temporal correlations during jhāna.** Detrended fluctuation analysis reveals that compared to mindfulness, jhāna is associated with reduced LRTC in the theta (C) (4–8 Hz; p = 0.050), (D) alpha (8–13 Hz; p = 0.023), and (E) beta (13–30 Hz; p = 0.001) band. In contrast, LRTC were significantly higher in the (F) low gamma band (30–45 Hz; p = 0.039), and trended higher in the (B) delta band (1–4 Hz; p = 0.150). (A) LRTC was not significantly different between jhāna and mindfulness when computed across the entire 1-45hz spectrum.

### *Jhāna increases the MMN compared to Mindfulness*

To complement the resting-state results, we examined whether jhāna alters auditory prediction error processing. After the silent EEG recording, participants continued meditating in their deepest stable state (jhāna or mindfulness) while passively listening to sequences of auditory stimuli that included a global oddball in 20% of cases, enabling us to probe pre-attentive sensory deviancy detection without severely disrupting the meditation.

As preregistered, we conducted paired t-tests on the mismatch negativity (MMN) amplitude at electrode Fz in two time windows: a time window in which the MMN typically peaks (0.1–0.25 s) and a later window (0.25–0.5 s). Neither test reached significance (p = 0.2406 and p = 0.0914, respectively). However, to better account for between-subject variability in brain anatomy and the scalp topography of the MMN, we repeated the analysis for an average of four frontocentral electrodes (Fz, Cz, FC1, FC2) for which the condition-averaged evoked response peaked, as is more commonly done in the literature (Duncan et al., 2009). This analysis revealed a significant increase in evoked response by oddball vs. standard stimuli during jhāna compared to mindfulness for the MMN (0.1–0.25s, p = 0.0370), but not for the later time window (0.25-0.5s, p = 0.1498). A cluster-based permutation test taking into account activity across all electrode sites further confirmed a significant enhancement of MMN in jhāna versus mindfulness in the early window (p = 0.0494), which was significant for the 185 to 229 ms past stimulus onset.

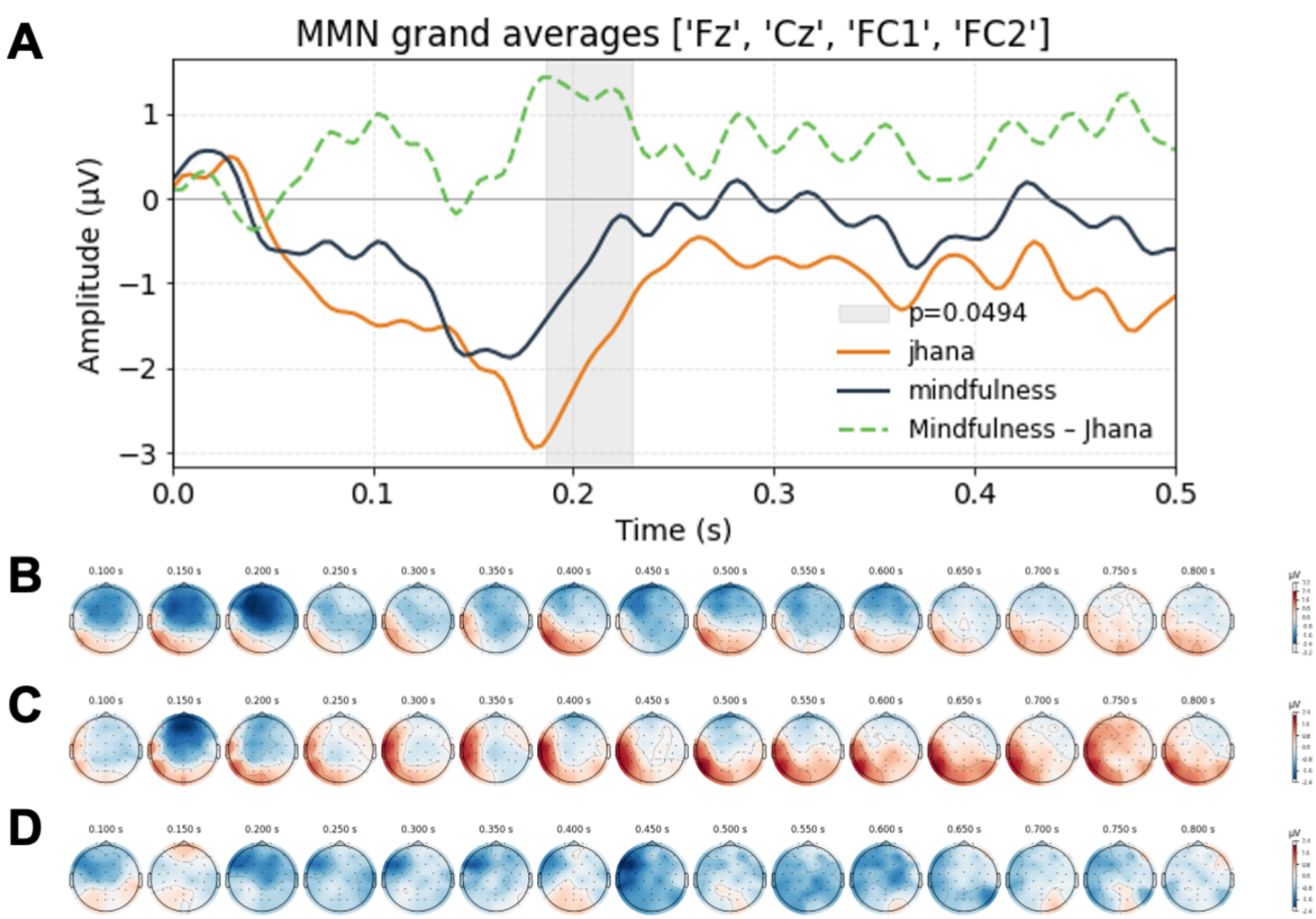

**Figure 4. Enhanced mismatch negativity during jhāna despite sensory withdrawal.** (A) Grand-average ERP at frontocentral electrodes showing the MMN response during jhāna (orange), mindfulness (blue), and the difference waveform (green). Despite phenomenological sensory fading, jhāna shows enhanced MMN ($p = 0.037$ for paired sample ttest, and $p=0.0494$ for cluster based permutation test). Topographic maps show (B) the jhāna MMN, (C) the mindfulness MMN, and (D) the difference wave (jhāna minus mindfulness) at 50-ms intervals from 100 to 800 ms after stimulus onset. Evoked responses are aligned to the onset of the final tone in the five-vowel tone sequence. For example, in the sequence A O A O **A**, $t = 0$ is defined as the onset of the final A (in bold).

## Discussion

The present study is the first to directly compare jhāna meditation to another active form of meditation, thereby isolating the neural mechanisms of meditative absorption. We found that jhāna practice, compared to a mindfulness of breath control condition, shifts brain dynamics to a regime that appears closer to criticality. These results indicate that brain criticality can be volitionally tuned through the self-regulation of attention. Compared to the mindfulness control, jhāna was characterized by increased signal diversity, reduced chaoticity, and an enhanced MMN response. Self-report ratings confirmed that absorptive depth distinguished jhāna from mindfulness, supporting our hypothesis that the state of absorption, rather than meditation practice per se, predicts proximity to criticality. Taken together, these findings suggest that jhāna meditation can reliably modulate brain dynamics without pharmacological intervention, and that brain criticality may serve as a neurophysiological marker of absorptive depth.

Our findings extend prior reports of increased signal diversity in jhāna relative to resting baseline control (Lieberman et al., 2025; Potash, van Mil, et al., 2025; Shinozuka et al., 2025) by showing that diversity increases more strongly in jhāna than in an active mindfulness control. This indicates that elevated signal diversity is not a general feature of meditation but is linked to the depth of absorption, thereby distinguishing jhāna from mindfulness practices. Importantly, in our framework signal diversity is not an isolated phenomenon but part of a broader dynamical shift: greater diversity, together with reduced chaoticity and enhanced MMN, reflects the movement of neural activity toward a critical regime. In this way, signal diversity provides a key index of—but does not exhaust—the broader reconfiguration of brain dynamics associated with criticality.

### *Jhāna Induces Volitional Criticality Through Structured Complexity*

Consistent with theoretical frameworks identifying criticality as the dynamical phase transition between order and chaos, compared to mindfulness, jhāna showed increases in multiple measures of signal diversity (Lempel–Ziv complexity, sample entropy, and Hjorth mobility), while simultaneously reducing signal chaoticity (Lyapunov exponents). In practical terms, this means that during jhāna, the brain explores a richer and less redundant repertoire of neural states (diversity increases), while those states remain dynamically constrained and do not diverge chaotically over time (chaoticity decreases). The coexistence of enhanced complexity and reduced chaoticity suggests has been described as a metastable regime with a broadened yet constrained repertoire of neural states (Toker et al., 2022). This aligns with recent definitions of "dynamically balanced criticality" in cortical networks, where neural systems exhibit flexible, transient responses without tipping into instability (Fosque et al., 2021).

The nature of this critical state becomes clearer when examining how the brain responds to unexpected sensory events during absorption. Jhānic states are classically described as involving deep stillness and withdrawal from sensory input. Consistent with this, our participants reported an experience of progressive sensory fading during jhāna, paralleled by slowed respiration and altered autonomic patterns. Yet this fading was not accompanied by the collapse of neural complexity seen in anesthesia or sleep. Despite consistent phenomenological reports of sensory withdrawal, during jhāna, response to the MMN was significantly enhanced, indicating increased sensitivity to deviations in sensory input. Increased responsivity to the MMN may suggest that rather than cortical isolation or sensory numbing, jhāna-induced states maintain, or even enhance, the brain's sensitivity to external perturbation, perhaps a reflection of the absorption depths unique to jhana. In line with this, several studies have reported an enhanced MMN during Focused Attention Meditation in expert meditators (Biederman et al., 2016; Fucci et al., 2018; Srinivasan & Baijal, 2007b). Importantly, by dissociating predictive deviancy detection from sensory adaptation, here we demonstrate that the enhanced MMN is specifically associated with a prediction error signal. However, another study did not find an effect of Focused Attention meditation on the MMN in expert meditators (Fucci et al., 2022). In that study, participants were asked to focus on a fixation cross in a lab setting as the Focused Attention meditation manipulation. The current study in contrast involved meditative absorption and was conducted during an ongoing retreat. This may have amplified effects on the MMN. That the MMN was present even in a deep absorption state aligns with the theory that sensory deviancy detection is a relatively automatic process in the awake brain. Furthermore, the jhāna state MMN response showed a central, left-lateralized frontal distribution, likely reflecting our use of linguistic stimuli (vowels) in the MMN (Näätänen et al., 1997).

Importantly, the increased responsivity to the MMN differentiates the neurophysiological signature of jhāna from psychedelics. Although both contexts are associated with increased level of brain criticality (Gervais et al., 2023; Toker et al., 2022) they differ in their response to the MMN: we observed increased sensitivity to the MMN during Jhāna, whereas psychedelics are generally associated with a blunting of the MMN (Erritzoe et al., 2024; Heekeren et al., 2008; Kometer et al., 2011, 2012, 2013). For example, ketamine, an NMDA receptor antagonist, consistently diminishes MMN amplitudes in healthy humans, closely paralleling the deficits seen in schizophrenia (Rosburg & Kreitschmann-Andermahr, 2016; Umbricht et al., 2002). Likewise, among classic serotonergic psychedelics, LSD reduces the MMN response in both the auditory (Heekeren et al., 2008) and visual (Murray et al., 2022) oddball paradigm. DMT also attenuates MMN, though less strongly than ketamine (Slater, 2020). Psilocybin has shown mixed effects: auditory studies reported little change (Bravermanová et al., 2018; Cavanna et al., 2022), whereas tactile paradigms revealed reduced mismatch responses to unexpected touch (Duerler et al., 2022). Taken together, these results suggest that while jhāna absorption enhances deviance detection, psychedelics typically dampen the brain’s surprise response.

### *Embodied Signatures of Critical Brain States*

Our peripheral physiological measures indicate that jhāna is not a passive or low-arousal state, but an actively regulated mode of embodied cognition. Compared to mindfulness, jhāna induced slower, more stable respiration and reduced variability in cardiac rhythms, suggesting coherent autonomic reorganization. Respiratory complexity increased while RR interval complexity decreased, pointing to simultaneous stabilization of cardiac output and dynamic adaptation of breathing patterns. These embodied changes may support sustained absorption by creating a stable physiological foundation that allows the brain to maintain a near-critical regime without being easily perturbed.

## Conclusions

Our findings provide evidence that jhāna meditation in experienced subjects reorganizes brain dynamics toward a structured, high-complexity regime. These results suggest that volitional, attention-based practice can reliably induce brain states approximating criticality without the aid of pharmacological interventions. Depth of absorption, rather than meditation practice per se, appears to determine the extent to which contemplative training moves neural dynamics toward criticality.

We treat jhāna and Metzinger's minimal phenomenal experience (MPE) as distinct yet convergent phenomena. Early fine-material jhānas retain an object (e.g., the breath nimitta) and a felt sense of absorption and unification, yet MPE is characterized by a near-absence of intentional content and loss of egoic self-awareness . Nevertheless, a central phenomenological characteristic of both jhāna states and MPEs is the quality of absorption, including a pronounced reduction of sensory perception. Because such progressive attenuation of phenomenal content is a central quality of minimal phenomenal experience (MPE), we propose that criticality may serve as a neurophysiological marker of absorption-related minimality. In this sense, criticality is not claimed as a universal signature of MPE, but as a candidate neurophysiological correlate of the absorptive dimension that brings consciousness closer to minimality.

While our research proposes that criticality is one neurophysiological marker of absorption and thus MPE, it is nevertheless important to acknowledge that MPEs as defined by Thomas Metzinger span a range of phenomenological characteristics of which there is some, but not a complete, overlap with the phenomenology of jhāna. Given this, we hope that future work will test whether criticality reliably indexes MPE-like experiences in other contemplative practices.

It is also important to acknowledge that the concept of criticality remains a rather novel construct in neurophysiology. Although well-established in physics, cognitive neuroscience has not yet converged on standard neural markers, and the reliability and interpretability of available indices such as complexity, chaoticity, and long-range temporal correlations remain under active debate. Early toolboxes provide promising methods, yet it is not clear which markers best capture criticality across brain states and contexts. While our results provided evidence for the validity of our chosen measures as a proxy for criticality, alternative approaches exist, including network-level metrics such as avalanche distributions or eigenvalue spectra as well as the 0-1 chaos test (Cocchi et al., 2017; Gervais et al., 2023; O'Byrne & Jerbi, 2022; Toker et al., 2024). Future work should pair phenomenological markers of MPE alongside multiple criticality measures and diverse contemplative practices to assess the generalizability of the link between experiential minimality and neural criticality. Taken together, this study demonstrates that meditative absorption provides a tractable, volitional route to states of the brain that approximate criticality.

## Methods

### *Participants*

Ten experienced lay practitioners from North America that have previously attained jhāna on retreat (7 female, 3 male; all right-handed; mean age = 61.3 years, SD = 12.1, range = 45–76) participated in the study during a 10-day residential retreat led by Shaila Catherine. Participants reported an average of 8,820 hours of lifetime meditation experience (SD = 9,090; range = 2,260hr to 32,173hr), including both retreat and at-home practice. All procedures were approved by the ethics committees of the Jewish General Hospital and the University of California, Berkeley. Informed consent was obtained from all participants.

### *Study Design and Procedure*

This study employed a within-subjects design conducted during the final four days of a 10-day in-person silent jhāna meditation retreat in North America. The retreat provided participants six full days to stabilize their attention and deepen their access to jhāna before data collection began. All participants were personally invited by the retreat teacher to ensure high-level expertise in the practice. Each of the final four days included a single 25-minute meditation session, alternating between two conditions: Mindfulness of Breathing and Jhāna Absorption. Condition order was randomized across days and counterbalanced across participants. Two participants meditated simultaneously in silence in the same room while EEG and physiological data were recorded using a 32-channel semi-dry EEG setup from bitbrain. A total of 40 sessions (10 participants × 4 sessions) were analyzed.

In the Mindfulness condition, participants were instructed to sustain attention on breath sensations—either at the nose or abdomen, according to personal preference—consistent with standard concentration-based mindfulness techniques. In the Jhāna condition, participants followed a structured attentional sequence that began by focusing attention on the occurrence of breath in the area of the nostrils. As concentration develops, physical sensations connected with the breath diminish and a visual sign (nimitta) emerges that replaces the physical breath as the object of focus. With concentration stabilized on the breath nimitta, practitioners entered the first jhāna and from there moved through the four fine-material jhānas. The study was limited to the fine-material jhānas as they were more reliably accessible to our participants. To track the progression through the four fine-material jhānas, participants were instructed to press a button each time they transitioned into a new jhāna stage. During mindfulness sessions, participants were instructed to press the button if they noticed a discrete or significant deepening of their meditative state.

Immediately following each session, participants answered short phenomenological questions confirming the meditative state practiced and the degree of sensory fading experienced. Semi-structured interviews were conducted post-retreat to further verify each participant's familiarity with jhāna stages and to confirm that all four absorption states had been reliably accessed during the study.

***MMN***

After every 25-minute meditation session in silence, participants continued in their deepest meditation for another 12 minutes during which we presented subjects with an auditory oddball task designed to elicit mismatch negativity (MMN). The auditory oddball task presented five vowels in a sequences like this "AOAOA," where the five tone sequences played each tone with 150ms stimulus-onset asynchrony. Thus, onset of the five tones occurred in a non randomized order at 0, 150ms, 300ms, 450ms, and 600ms relative to the sequence start. Sequences were separated by a jittered inter-trial interval (ITI) between 1350ms and 1650ms with 50ms intervals, yielding 1.95 to 2.25s onset-to-onset spacing across consecutive sequences. The global oddball condition introduced a deviation in the final tone of the sequence, resulting in "AOAO**O**"/"OAOA**A**" instead of the standard sequence "AOAO**A**"/"OAOA**O**". To familiarize participants with the task, 20 preparatory trials were conducted without any oddballs. This was followed by 350 sequences, of which 80% were standard sequences and 20% were oddballs, randomly intermixed. These vowels were generated by the VowelEditor functionality of the computer program Praat (Boersma & Weenink, 2001).

We chose to employ vowel sounds rather than the more typical pure tones, following the approach of Strauss and colleagues (2015) who examined oddball responses during sleep. Using vowels also provided tighter control over potential low-level frequency confounds between stimuli. Across the different blocks of the experiment, each vowel was presented as a standard and as a deviant to ensure balanced exposure.

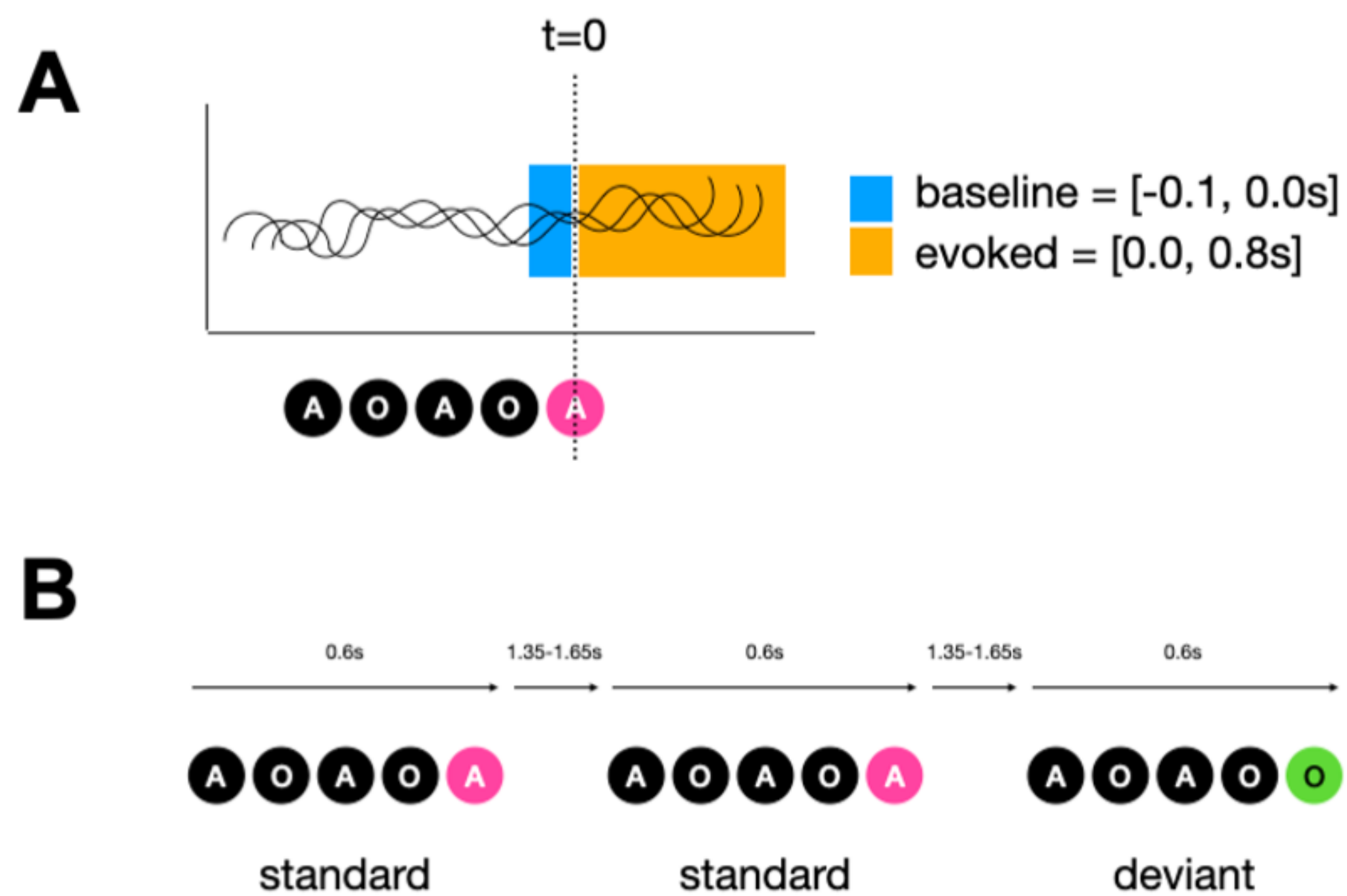

**Figure 5.** Auditory oddball paradigm used to elicit mismatch negativity (MMN). (A) Example trial structure. Each sequence consisted of five vowels presented with 150 ms stimulus-onset asynchrony, yielding onsets at 0, 150, 300, 450, and 600 ms relative to sequence start. The final tone in the sequence served as the critical event (t = 0) for EEG analysis, with a baseline window of –100 to 0 ms and an evoked window of 0 to 800 ms. (B) Illustration of standard and deviant sequences. Standard sequences ended with a repetition of the alternating pattern (e.g., "AOAOA"), whereas deviant sequences violated this pattern at the final position (e.g., "AOAOO"). Sequences were separated by a jittered inter-trial interval of 1350–1650 ms (50 ms steps), producing ~1.95–2.25 s onset-to-onset spacing across trials. Each block began with 20 standard-only preparatory sequences, followed by 350 mixed trials (80% standards, 20% deviants).

***Data collection and preprocessing***

EEG and physiological signals were recorded during each 40-minute meditation session in a quiet, secluded setting to minimize environmental distractions. Recordings were performed using a 32-channel mobile EEG system (Bitbrain) with electrodes arranged in the international 10–20 configuration, referenced to the left ear and grounded frontally. EEG data were sampled at 256 Hz. In addition to EEG, we collected respiration (via chest belt), electrocardiography (ECG; lead II configuration), and electrodermal activity (EDA; via electrodes on the small and middle fingers of the left hand) to assess autonomic physiology. EEG data were first manually cleaned through visual inspection. Segments with excessive noise or non-physiological artifacts were marked and excluded, and persistently noisy channels were removed. This manual cleaning was prioritized over automated ICA-based denoising, as participants maintained eyes-closed stillness throughout, and few stereotypical artifact components (e.g., eye blinks or muscle bursts) were evident given the controlled retreat setting. For most participants, no clear ICA components were associated with artifacts beyond those already excluded manually, and ICA was therefore not applied. Following manual cleaning, EEG data were processed in two parallel streams. For epoch-based analyses, the remaining clean data were segmented into non-overlapping 10-second epochs. For continuous time-series analyses, manually marked bad segments were removed from the raw data, and all remaining clean segments of at least 20 seconds in duration were concatenated to form continuous time series.

### *Feature extraction*

Following preprocessing, EEG data were analyzed in two complementary streams: 10-second non-overlapping epochs for complexity and spectral measures, and continuous concatenated segments for avalanche and fractal analyses. All analyses were performed using custom Python scripts leveraging MNE (Gramfort et al., 2013), NeuroKit2 (Makowski et al., 2021), AntroPy (Vallat, 2022), and FOOOF (Donoghue et al., 2020), with multiprocessing for efficient computation. The scripts will be made publicly available on GitHub upon publication and are available to reviewers upon request.

### *Signal Diversity and Complexity Measures*

To characterize neural signal diversity and complexity, we computed a suite of time-domain and information-theoretic features from artifact-free EEG epochs filtered between 0.5 and 40 Hz. Lempel–Ziv complexity (LZC) was calculated by first binarizing each channel's time series based on the median amplitude and quantifying the number of unique binary substrings relative to the signal's length, yielding a normalized estimate of compressibility. Sample entropy was computed by assessing the likelihood that sequences of two consecutive data points (embedding dimension m = 2) remain similar at the next point within a tolerance of 20% of the signal's standard deviation. Permutation entropy was derived by converting each time series into a sequence of ordinal patterns with an embedding dimension of 3 and evaluating the Shannon entropy of the resulting symbol distribution, with normalized values. Spectral entropy was calculated from the normalized power spectral density using Welch's method, reflecting the flatness of the frequency distribution across the 0.5–45 Hz band. Hjorth mobility was estimated as the standard deviation of the signal's first derivative divided by the standard deviation of the original signal, while Hjorth complexity captured the change in frequency by comparing the mobility of the derivative to that of the signal itself.

Fractal scaling properties were captured using two methods: the Hurst exponent was computed by measuring the scaling of rescaled range statistics across exponentially spaced window sizes from 1 to 3 seconds, and detrended fluctuation analysis (DFA) estimated the slope of fluctuation amplitudes as a function of scale after linear detrending. To probe dynamical sensitivity, the largest Lyapunov exponent was estimated using NeuroKit2's (Makowski et al., 2021) implementation of the Rosenstein (1993) algorithm for short time series, wherein nearby trajectories in reconstructed phase space were tracked to measure their average exponential divergence, quantifying chaoticity. The correlation dimension was computed by reconstructing the phase space of each signal and quantifying how the number of point pairs within a given distance scales with that distance, estimating the signal's effective dimensionality. Finally, multiscale entropy was computed by coarse-graining each time series at multiple scales and calculating the average sample entropy across these scales, providing a composite measure of complexity across temporal resolutions. All features were computed per epoch and channel, then averaged across non-interpolated channels to obtain subject- and condition-level summary statistics.

### ***Spectral Measures and 1/f Slope***

Power spectral density (PSD) was estimated using Welch's method, in which EEG time series were segmented into overlapping windows and tapered with a Hamming window before computing the averaged periodogram, yielding robust estimates of frequency-domain power. To characterize scale-free dynamics, we modeled the PSD using the FOOOF algorithm, which decomposes the spectrum into an aperiodic (1/f-like) component and superimposed oscillatory peaks (Donoghue et al., 2020). The slope and intercept of the aperiodic component were extracted by fitting a linear model to the log–log-transformed PSD within a frequency range of 0.5–45 Hz, excluding peak-like deviations to isolate the broadband background structure. These parameters were calculated globally from the mean PSD across channels, as well as independently for each electrode to assess spatial heterogeneity. To isolate genuine oscillatory power, the fitted aperiodic component was subtracted from each channel's PSD, yielding a corrected spectrum. Band-limited power was then computed by averaging the residual power within canonical frequency bands (delta: 0.5–4 Hz; theta: 4–8 Hz; alpha: 8–12 Hz; beta: 12–30 Hz; gamma: 30–45 Hz), allowing us to assess frequency-specific effects while controlling for broadband shifts.

***MMN***

The same preprocessing as described above for the resting state data was applied to the recordings of the hierarchical auditory-oddball task. Cleaned EEG was segmented into epochs time-locked to the onset of the fifth tone in each five-vowel sequence ($t = 0$). Epochs spanned −0.10 to 0.80 s, were baseline-corrected to the pre-stimulus window from −100 to 0 ms, and low-pass filtered at 40 Hz. For each participant and day, we computed evoked responses separately for standards and deviants with the MMN being defined as the difference wave (deviant minus standard). Day-level averages were then combined with equal weights to yield one evoked response per condition (standard, deviant, MMN) and meditation state (jhāna, mindfulness) for each participant. Analyses focused on the fronto-central region and were carried out both at electrode Fz alone and at the average of Fz, Cz, FC1, and FC2 to reduce topographical variability.

Following common MMN practice, we defined two analysis windows: an early window from 100 to 250 ms post-stimulus onset, covering the canonical MMN peak, and a later window from 250 to 500 ms to test for sustained or secondary effects. For each participant we averaged the amplitude within each window and compared jhāna and mindfulness conditions using paired tests.

To complement the windowed analyses and to control for family-wise error across time, we additionally performed a cluster-based permutation test on the averaged time series from the four fronto-central electrodes (Fz, Cz, FC1, FC2) within a pre-specified window of 100–250 ms post-stimulus onset. In this test, contiguous time samples showing consistent condition differences were grouped into clusters, and the sum of their t-values served as the cluster statistic. A null distribution was generated by randomly permuting condition labels within participants 100,000 times, with the maximum cluster statistic from each permutation retained. Observed clusters were then assigned p-values based on their position within this distribution, providing rigorous control for multiple comparisons. We used a two-sided test and, rather than applying a fixed cluster-forming threshold (e.g., $t = 2.6$), adopted a threshold-free approach in which the permutation procedure itself identified clusters, yielding a more data-driven estimate of significance.

***Statistical analysis***

Prior to statistical testing, outliers were removed using an interquartile-range (IQR) rule: values less than $Q1 – 3 \times IQR$ or greater than $Q3 + 3 \times IQR$ were excluded (IQR multiplier = 3.0). Outlier detection was performed on the combined mindfulness + jhāna distribution for each measure, and the corresponding rows were dropped from all subsequent analyses for that measure. Plots indicating removed points are provided in the Appendix.

To leverage all trials/epochs while accounting for repeated measures, we fit linear mixed-effects models with random intercepts per subject, using restricted maximum likelihood (REML). Unless otherwise noted, the model specification was: DV ~ condition + day + (1 | subject) with mindfulness as the reference level for condition and day treated as a categorical factor. For clarity in figures, models used a z-scored dependent variable (standardized across included observations) so that beta coefficients are in SD units.

All tests are two-tailed with alpha = 0.05. Across different measures/frequency bands we report uncorrected p values (unless noted); convergent results across theoretically linked metrics (e.g., LZC, sample/permutation entropy, Hjorth mobility, 1/f slope) mitigate—but do not eliminate—the multiple-testing burden.

For each mixed model we report a pseudo-$R^2$ computed as the coefficient of determination between observed outcomes and model fitted values (fixed + random effects). This is not Nakagawa's marginal/conditional $R^2$; it quantifies variance explained by the full fitted values. Because random intercepts account for stable between-subject differences, these pseudo-$R^2$ values are typically higher than conventional marginal $R^2$ and should be interpreted as overall goodness of fit rather than variance explained by condition alone.

Day effects were assessed with a within-subject repeated-measures ANOVA across the four days (computed separately for the jhāna and mindfulness condition), and we report partial eta-squared ($\eta^2$) as the effect size. Pairwise day comparisons were adjusted using Holm correction.

Analyses were performed in Python using statsmodels (Seabold & Perktold, 2010), scipy (Virtanen et al., 2020), and scikit-learn (Pedregosa et al., 2011). figures were generated with matplotlib (Hunter, 2007) and seaborn (Waskom, 2021).

## Appendix

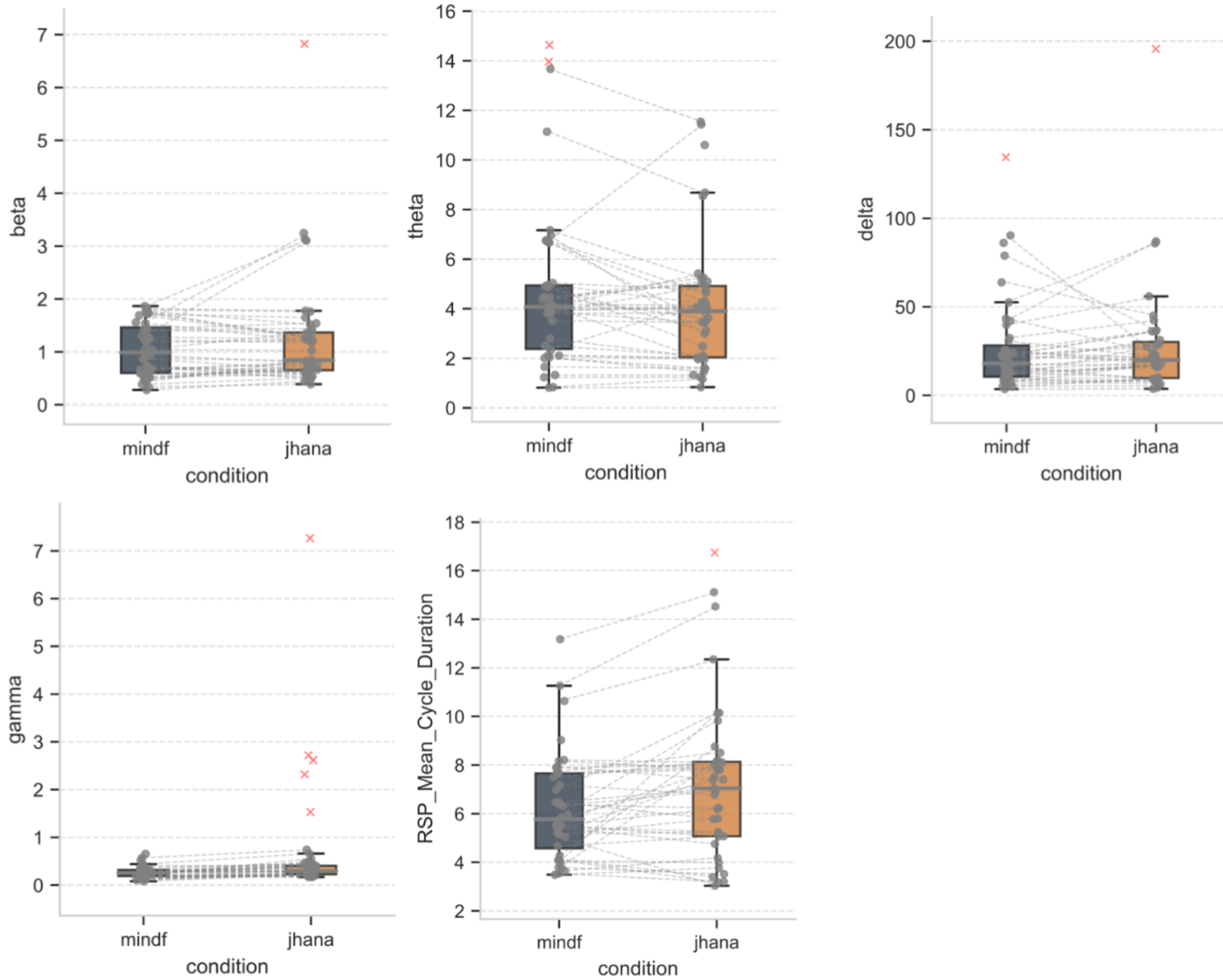

**Appendix A**. Outlier detection using the interquartile range (IQR) method. Values below Q1 – 3 × IQR or above Q3 + 3 × IQR were excluded prior to statistical testing. Excluded outliers are shown as red × markers.